\documentclass[twocolumn,showpacs,aps,prb,superscriptaddress,floatfix]{revtex4}
\usepackage{graphicx}% Include figure files
\usepackage{dcolumn}% Align table columns on decimal point
\usepackage{bm}% bold math
\begin{document}
\preprint{}
%
%\draft

\title{Critical properties of the unconventional spin-Peierls system TiOBr}

\author{J.P. Clancy}
\affiliation{Department of Physics and Astronomy, McMaster University,
Hamilton, Ontario, L8S 4M1, Canada}

\author{B.D. Gaulin}
\affiliation{Department of Physics and Astronomy, McMaster University,
Hamilton, Ontario, L8S 4M1, Canada}
\affiliation{Canadian Institute for Advanced Research, 180 Dundas St. W.,
Toronto, Ontario, M5G 1Z8, Canada}

\author{F.C. Chou}
\affiliation{Center for Condensed Matter Sciences, National Taiwan University, Taipei 106, Taiwan.}

\begin{abstract}
We have performed detailed x-ray scattering measurements on single crystals of the spin-Peierls compound TiOBr in order to study the critical properties of the transition between the incommensurate spin-Peierls state and the paramagnetic state at T$_{C2}$ $\sim$ 48 K.  We have determined a value of the critical exponent $\beta$ which is consistent with the conventional 3D universality classes, in contrast with earlier results reported for TiOBr and TiOCl.  Using a simple power law fit function we demonstrate that the asymptotic critical regime in TiOBr is quite narrow, and obtain a value of $\beta_{asy}$ = 0.32 $\pm$ 0.03 in the asymptotic limit.  A power law fit function which includes the first order correction-to-scaling confluent singularity term can be used to account for data outside the asymptotic regime, yielding a more robust value of $\beta_{avg}$ = 0.39 $\pm$ 0.05.  We observe no evidence of commensurate fluctuations above T$_{C1}$ in TiOBr, unlike its isostructural sister compound TiOCl.  In addition, we find that the incommensurate structure between T$_{C1}$ and T$_{C2}$ is shifted in {\bf Q}-space relative to the commensurate structure below T$_{C1}$. 

\end{abstract}

\maketitle
% \narrowtext
% \twocolumn
% body of paper here

\section{Introduction}

The physics of low dimensional quantum magnets is a very interesting and topical area of research.  These systems exhibit a broad range of novel physical properties and exotic quantum ground states, ranging from spin-Peierls and Haldane gap behavior in quasi-one-dimensional systems to high temperature superconductivity in quasi-two-dimensional systems.  Spin-Peierls systems in particular have undergone a recent renaissance due to the discovery of two new, and somewhat unconventional, inorganic spin-Peierls compounds, the titanium oxyhalides TiOBr and TiOCl.

Spin-Peierls behavior arises in quasi-one-dimensional (1D) systems which possess a combination of spin 1/2 magnetic moments, short-range antiferromagnetic interactions, and strong magneto-elastic coupling.  At low temperatures these systems undergo a spin-Peierls phase transition which is characterized by dimerization of the lattice and the development of a non-magnetic singlet ground state.  As the presence of interchain interactions typically drives quasi-1D systems towards long-range magnetic order, only a select family of compounds with particularly strong spin-phonon coupling have been found to exhibit a spin-Peierls transition.  The first experimental realization of the spin-Peierls transition was discovered in organic charge transfer compounds, such as TTF-CuBDT\cite{Spin-Peierls} and MEM-(TCNQ)$_{2}$\cite{MEM}.  It was considerably later that spin-Peierls behavior was observed in CuGeO$_{3}$, the first inorganic spin-Peierls compound (T$_{SP}$ $\sim$ 14 K)\cite{CuGeO3}.

TiOBr and TiOCl, the recently discovered inorganic spin-Peierls compounds based on Ti$^{3+}$, are often referred to as ``unconventional" spin-Peierls systems.  Unlike standard spin-Peierls systems, which undergo a single continuous phase transition at T$_{SP}$, the titanium oxyhalides undergo two successive phase transitions upon warming - a first-order transition at T$_{C1}$ between commensurate and incommensurate spin-Peierls states, and a higher-order transition at T$_{C2}$ between the incommensurate spin-Peierls state and a disordered pseudogap state which extends up to T* $\sim$ 135 K\cite{Imai}.  In addition, the titanium oxyhalides are unique among spin-Peierls compounds because of their unusually high transition temperatures (T$_{C1}$/T$_{C2}$ $\sim$ 65 K/92 K in TiOCl\cite{Seidel, Imai, Krimmel, Ruckamp, Hemberger} and 27 K/48 K in TiOBr\cite{Sasaki_JPSJ, Lemmens, Smaalen}) and surprisingly large singlet-triplet energy gap (E$_{g}$ $\sim$ 430 to 440 K in TiOCl\cite{Imai, Baker} and 150 K in TiOBr\cite{Sasaki_JPSJ}).  The origins and properties of the incommensurate spin-Peierls state between T$_{C1}$ and T$_{C2}$ have attracted particular attention to these compounds.

Both TiOBr and TiOCl crystallize into a FeOCl-type structure, with an orthorhombic space group of \textit{Pmmn} at room temperature\cite{Schafer, Schnering}.  This structure consists of buckled Ti-O bilayers in the \textit{ab}-plane, which are separated by double layers of either Br$^{-}$ or Cl$^{-}$ ions and stacked along the \textit{c}-axis.  The additional reduction of effective dimensionality within the Ti-O bilayers, from two to one, is believed to arise from ordering of the Ti$^{3+}$ orbital degrees of freedom\cite{Seidel}.  Spin-Peierls dimerization occurs along the crystallographic \textit{b}-axis, and is driven by the direct exchange of Ti \textit{d}$_{xy}$ orbitals\cite{Seidel}.  High temperature susceptibility measurements suggest that both TiOBr and TiOCl can be fit to a S = 1/2 Heisenberg chain model, with nearest-neighbor exchange couplings of $\sim$ 660 K and 360 K respectively\cite{Seidel, Kato, Sasaki_JPSJ}.

In general, TiOBr has attracted considerably less attention than its sister compound TiOCl.  This is largely due to the fact that it is more difficult to grow single crystals of TiOBr than TiOCl, although it is also more challenging to work with TiOBr because it is extremely hygroscopic (even more so than its isostructural counterpart) and has substantially lower transition temperatures.  All measurements to date seem to suggest that apart from a slight difference in energy scales, the two compounds behave almost identically - they have matching low temperature phase diagrams, and appear to be isostructural in each of the major phases.  However, there are subtle differences between the lattice constants of the two materials, perhaps most notably in the ratio of the \textit{a} and \textit{b} lattice constants which describe the Ti-O bilayers.  While the room temperature lattice parameters reported for TiOBr are \textit{a} = 3.78 {\AA}, \textit{b} = 3.49 {\AA}, and \textit{c} = 8.53 {\AA}\cite{Sasaki_JPSJ}, those of TiOCl are \textit{a} = 3.79 {\AA}, \textit{b} = 3.38 {\AA}, and \textit{c} = 8.03 {\AA}\cite{Schafer}.  Thus, while the \textit{a}-axis (i.e. interchain) spacing of the two compounds is almost identical, the \textit{b}-axis (i.e. intrachain) spacing is $\sim$ 3 to 4\% smaller in TiOCl.  As a result, both the relative and absolute strength of the intrachain interactions are expected to be greater in TiOCl, giving rise to physical properties which are more strongly one-dimensional in nature.  This appears to be reflected in both the magnitude of the nearest-neighbor exchange coupling and the size of the singlet-triplet energy gap.

The critical behavior of spin-Peierls systems is a subject which has generated considerable interest and confusion over the past thirty years.  Widely varying and often contradictory critical exponents have been reported for a number of spin-Peierls compounds.  Early studies of CuGeO$_{3}$, the first inorganic spin-Peierls system, led to conflicting claims of tricritical\cite{Harris_94, Fujita, StPaul}, three-dimensional\cite{Harris_95, Winkelmann, Lorenz}, and mean-field-like\cite{Birgeneau} critical behavior.  Subsequent work\cite{Lumsden_pure} has revealed that much of this discrepancy can be attributed to the narrowness of the asymptotic critical region, and it has been shown that the system belongs to a conventional 3D universality class.  Similarly, while the organic compounds MEM-(TCNQ)$_2$ and TTF-CuBDT were initially believed to exhibit mean-field-like critical behavior\cite{MEM_crit, TTF_crit}, more detailed measurements\cite{Lumsden_MEM} suggest that these compounds correspond to 3D universality classes as well.
  
The critical behavior of the recently discovered Ti-based spin-Peierls compounds, TiOBr and TiOCl, has been less extensively studied but appears similarly enigmatic.  To date there has been only one report of critical exponents for these systems, which placed the value of $\beta$ for the continuous transition at T$_{C2}$ (between the disordered pseudogap state and the incommensurate spin-Peierls state) between 0.10 $\pm$ 0.04 and 0.26 $\pm$ 0.09 in TiOBr\cite{Sasaki_condmat}.  Similar values of $\beta$ = 0.15\cite{Abel} and $\beta$ = 0.25\cite{Smaalen} can be extracted from other x-ray diffraction measurements on TiOCl and TiOBr, although these results have not been explicitly quoted as critical exponents by the authors.  This range of  $\beta$ values suggests some form of either quasi-2D or tricritical behavior at T$_{C2}$, a result which is somewhat surprising for a phase transition which is known to be driven by inherently three-dimensional spin-phonon interactions.  For two compounds which have already been shown to exhibit behavior which significantly differs from the standard spin-Peierls scenario, this poses a very interesting question - is the critical behavior of these compounds truly distinct from that of CuGeO$_{3}$ and the organic spin-Peierls systems or is 3D universality one of the fundamental properties of any spin-Peierls transition?  The need for a detailed study of the critical behavior of these systems has been further underlined by a recent claim based on symmetry analysis\cite{Schonleber_crackpot} that the transition at T$_{C2}$ is intrinsically first order in nature, despite the rather substantial weight of experimental evidence to the contrary\cite{Seidel, Imai, Ruckamp, Hemberger, Shaz, Smaalen, Abel, Sasaki_condmat}.

In this paper we report the first detailed study of the critical properties of TiOBr.  Using an approach which incorporates both standard power law fits and modified power law fits which include a first order correction-to-scaling term, we find that the transition at T$_{C2}$ can be well described by conventional 3D universality classes.  We offer a qualitative comparison which suggests that TiOCl also corresponds to a similar universality class, and discuss the potential change in criticality which results when TiOCl is doped with non-magnetic Sc ions.  Our results indicate that the asymptotic critical regime in TiOBr is quite narrow, providing a natural explanation for the anomalously low values of $\beta$ which have previously been reported in the literature.  The ease with which our data can be fit to models of a continuous phase transition clearly contradicts arguments for the first order nature of the spin-Peierls transition at T$_{C2}$\cite{Schonleber_crackpot}.  In addition, we offer evidence that commensurate fluctuations, which inhabit the incommensurate spin-Peierls and pseudogap phases of TiOCl, are absent in TiOBr.  We propose that this may reflect the weaker one-dimensionality of the system, and the greater importance of interchain interactions.  Finally, we observe that the incommensurate and commensurate spin-Peierls states in TiOBr appear to be shifted in {\bf Q}-space with respect to each other.  This may be an indication that the structure of the incommensurate spin-Peierls state is more complex than previously assumed. 

\section{Experimental Details}

Single crystal samples of TiOBr were prepared using the chemical vapor transport method.  A more detailed description of the crystal growth procedure has been reported elsewhere\cite{Lemmens}.  The dimensions of the sample used in this experiment were approximately 4.0 $\times$ 3.0 $\times$ 0.5 mm.  The sample was mounted on the cold finger of a closed cycle refrigerator and aligned within a Huber four circle diffractometer.  This sample environment provided a base temperature of $\sim$ 7 K, with temperature stability of better than $\pm$ 0.005 K.  X-ray scattering measurements were performed using Cu-K$\alpha$ radiation ($\lambda$ = 1.54 {\AA}) produced by an 18 kW rotating anode x-ray source with a vertically focussing pyrolytic graphite monochromator.

X-ray scattering scans were performed through several equivalent commensurate ordering wave vectors of the form (H, K+1/2, L), where superlattice Bragg peaks are expected to arise as the result of spin-Peierls dimerization.  Due to the thickness of our sample and the relatively short penetration depth of the wavelength being employed ($\sim$ 20 {$\mu$}m), these measurements were largely restricted to reflection geometry.  Ultimately, our measurements focused around the (0, 0.5, 6) position in reciprocal space, as it was here that the intensity of the superlattice Bragg peaks was found to be strongest.  Scans were carried out along the H, K, and L directions in reciprocal space in order to characterize all of the relevant commensurate and incommensurate superlattice peaks.  These measurements were repeated at several temperatures within each of the four major regions of the TiOBr phase diagram - below T$_{C1}$ in the commensurate spin-Peierls phase, between T$_{C1}$ and T$_{C2}$ in the incommensurate spin-Peierls phase, between T$_{C2}$ and T* in the pseudogap phase, and above T* in the paramagnetic phase.

Once the general features of the scattering at the superlattice peak positions had been well characterized, a more detailed study of the incommensurate scattering at (H $\pm$ $\delta_{H}$, K+1/2, L) was performed close to T$_{C2}$ in order to investigate the critical behavior of the system.  In particular, a series of scans were taken through the incommensurate ordering wave vector at (-$\delta_{H}$, 0.5, 6) to follow the temperature evolution of the incommensurate scattering intensity.  To ensure that these measurements captured all of the relevant incommensurate scattering, two long scans were performed at every temperature - one scan in the H-direction to locate the center of the incommensurate satellite peak, followed by a second scan in the K-direction to obtain a measure of the integrated scattering intensity.  Measurements were performed over a range of temperatures extending from $\sim$ 11 K below the phase transition to $\sim$ 3 K above the transition.  The size of the temperature steps varied from as large as 0.25 K (more than 6 K away from T$_{C2}$) to as small as 0.05 K (within 1 K of T$_{C2}$).  In order to check for the presence of thermal hysteresis at T$_{C2}$, two separate data sets were collected while warming and cooling through the transition temperature.  As no observable difference could be found between the measurements obtained on warming and cooling, both sets of measurements will be treated as a single data set in the analysis which follows in Section III.A.

\begin{figure}
\includegraphics{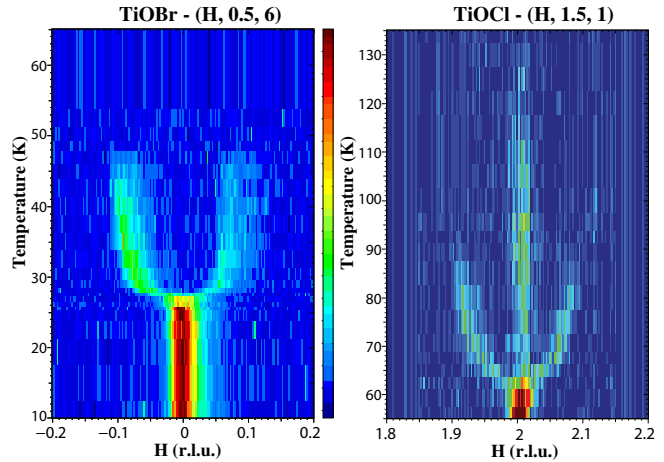}
\caption{(Color online) Color contour maps of x-ray scattering intensity as a function of temperature in TiOBr (left) and TiOCl (right).  These maps are composed of H-scans of the form (H, 0.5, 6) and (H, 1.5, 1) respectively.  A high temperature (150 K) background data set has been subtracted from the TiOCl data in order to eliminate higher order wavelength contamination. }
\end{figure}

\section{Results and Discussion}

\subsection{Critical Behavior of TiOBr}

Figure 1 shows two color contour plots, consisting of H-scans collected for TiOBr (left panel) and TiOCl (right panel).  Each of these maps is composed of a series of H-scans performed through one of the superlattice peak positions - the (0, 0.5, 6) position in the case of TiOBr, and the (2, 1.5, 1) position in the case of TiOCl.  Note that in the case of TiOCl a high temperature background data set, collected at T = 150 K, has been subtracted from each scan.  This correction is necessary in order to eliminate weak $\lambda$/2 contamination present in the incident x-ray beam, which can give rise to Bragg scattering at the superlattice peak positions.  The $\lambda$/2 scattering at (2, 1.5, 1) in TiOCl is comparable in strength to the weak $\lambda$ scattering in the pseudogap phase ($\sim$ 0.1 counts/s), but it can be eliminated with a simple background subtraction due to its lack of temperature dependence.  No higher order wavelength contamination was observed at the (0, 0.5, 6) position in TiOBr, so a corresponding high temperature background subtraction has not been performed.

The color contour maps in Fig. 1 offer a qualitative summary of the main features of the low temperature phase diagram of the titanium oxyhalides.  Below T$_{C1}$ there is a dimerized, commensurate two-fold superstructure, identified by a single superlattice peak at (H, K+1/2, L).  Between T$_{C1}$ and T$_{C2}$ two incommensurate satellite peaks appear at the (H $\pm$ $\delta_{H}$, K+1/2, L) positions, signifying the development of an incommensurately modulated version of the dimerized structure.  The primary component of the incommensurate modulation lies along the H-direction in reciprocal space, or the \textit{a}-axis in real space, which means that it runs perpendicular to the dimerized chains of Ti atoms in the spin-Peierls state.  Note that additional commensurate scattering, which has been attributed to commensurate fluctuations\cite{Clancy_pure}, is visible between T$_{C1}$ and T* in TiOCl but not in TiOBr.  This result will be discussed in greater detail in section III.C.  For the present time it is worth noting that, if anything, the physical picture for TiOBr is simpler than that of its isostructural counterpart.

\begin{figure}
\includegraphics{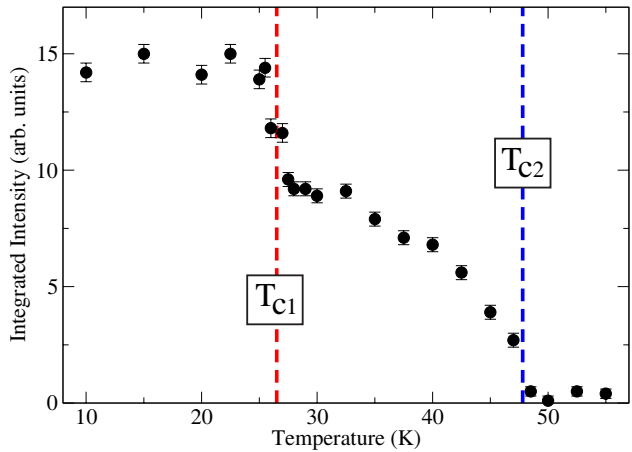}
\caption{(Color online) The temperature dependence of the integrated scattering intensity of the superlattice peaks observed at (0, 0.5, 6) [below T$_{C1}$, in the commensurate SP phase] and ($\pm \delta_{H}$, 0.5, 6) [above T$_{C1}$, in the incommensurate SP phase] in TiOBr.  The discontinuous drop in scattering intensity at T$_{C1}$ $\sim$ 27 K is indicative of a first order phase transition, while the gradual decrease towards T$_{C2}$ is consistent with a continuous phase transition. }
\end{figure}

TiOBr is known to undergo two successive phase transitions at low temperatures - a continuous transition into an incommensurate spin-Peierls state at T$_{C2}$ $\sim$ 48 K, followed by a first order transition into a commensurate spin-Peierls state at T$_{C1}$ $\sim$ 27 K.  These two transitions are clearly illustrated in Fig. 2, which shows the temperature dependence of the integrated scattering intensity in the vicinity of the (0, 0.5, 6) superlattice peak position.  Below T$_{C1}$ the scattering intensity arises from the commensurate superlattice peak at (0, 0.5, 6), while between T$_{C1}$ and T$_{C2}$ it originates from the incommensurate satellite peaks at ($\pm$ $\delta_{H}$, 0.5, 6).  All of the scattering intensities provided in Fig. 2 were obtained by integrating over H-scans of the form shown in Fig. 1.  It should be noted that the intensity of the scattering which results from spin-Peierls dimerization is proportional to the square of the relevant atomic displacements, and hence can be directly related to the square of the order parameters associated with the phase transitions at T$_{C1}$ and T$_{C2}$.  For this reason, the discontinuous jump in scattering intensity at T$_{C1}$ is evidence for a first order phase transition, while the gradual decrease towards T$_{C2}$ is indicative of a continuous phase transition.  Both of these results are consistent with the generally accepted nature of the transitions at T$_{C1}$ and T$_{C2}$.

\begin{figure}
\includegraphics{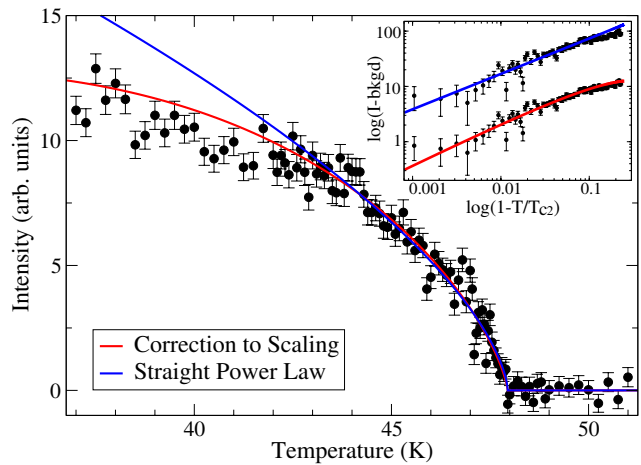}
\caption{(Color online) Fits to the integrated intensity of the incommensurate peak at (-$\delta_{H}$, 0.5, 6) as a function of temperature. These fits were performed using all data collected above T$_{i}$ = 43 K, employing both a straight power law fit function (blue, $\beta_{asy}$ = 0.32) and a power law including the first order correction-to-scaling confluent singularity term (red, $\beta_{avg}$ = 0.39).  In the inset the quality of these fits is shown on a logarithmic scale. }
\end{figure}

Since the intensity of the dimerization scattering is proportional to the square of the order parameter, it is possible to extract a value for the critical exponent $\beta$ from our x-ray diffraction data.  As only the phase transition at T$_{C2}$ is continuous, we will focus our attention on the critical behavior associated with the transition between the incommensurate spin-Peierls phase and the disordered pseudogap phase.  In this case, the order parameter for the phase transition will be related to the intensity of the incommensurate satellite peaks which appear at the (H $\pm$ $\delta_{H}$, K + 1/2, L) wave-vectors.  The integrated scattering intensity of the (-$\delta_{H}$, 0.5, 6) incommensurate peak, which was found to be the strongest of the experimentally accessible incommensurate peaks, was carefully determined by performing a series of H and K-scans as described in Section II.  The integrated scattering intensity was then fit to two different functional forms, a straight power law (Eq. 1) and a modified power law which includes the first order correction-to-scaling confluent singularity term (Eq. 2).

\begin{eqnarray}
I(t) = A t^{2 \beta} + background
\end{eqnarray}
\begin{eqnarray}
I(t) = A t^{2 \beta} \times \left(1 + B t^{\Delta}\right) + background
\end{eqnarray} 

Here $t = (T-T_{C2})/T_{C2}$ is the reduced temperature, while the value of the exponent $\Delta$ has been determined to be $\sim$ 0.5\cite{Aharony, Guillou}.  Representative fits to the integrated intensity data using Eqs. 1 and 2 are provided in Fig. 3.  The choice of two fit functions was motivated by previous studies on the critical behavior of CuGeO$_{3}$\cite{Lumsden_pure} which showed that the asymptotic critical region was extremely narrow, extending less than 0.5 K (or $\sim$ 4\%) below T$_{SP}$.  In this region the length scale associated with fluctuations in the order parameter is greater than any of the other relevant length scales that occur in the system.  Within the limit of the asymptotic critical regime it is expected that the order parameter should be accurately described by the simple power law provided in Eq. 1.  Farther away from the transition temperature however, outside the asymptotic critical regime, other length scales in the system become significant and the behavior of the order parameter begins to deviate from this ideal form.  It is this effect which was responsible for the wide variation between the initially reported critical exponents for CuGeO$_{3}$, and we propose that the same holds true for TiOBr.  In order to account for deviations from standard power law behavior farther away from T$_{C2}$, we can modify our original fit function to include the first order correction-to-scaling confluent singularity term, as shown in Eq. 2.  It should be noted that this correction is only the first term in a larger series expansion which describes the behavior of the order parameter beyond the limits of the asymptotic critical regime\cite{Aharony}.

\begin{figure}
\includegraphics{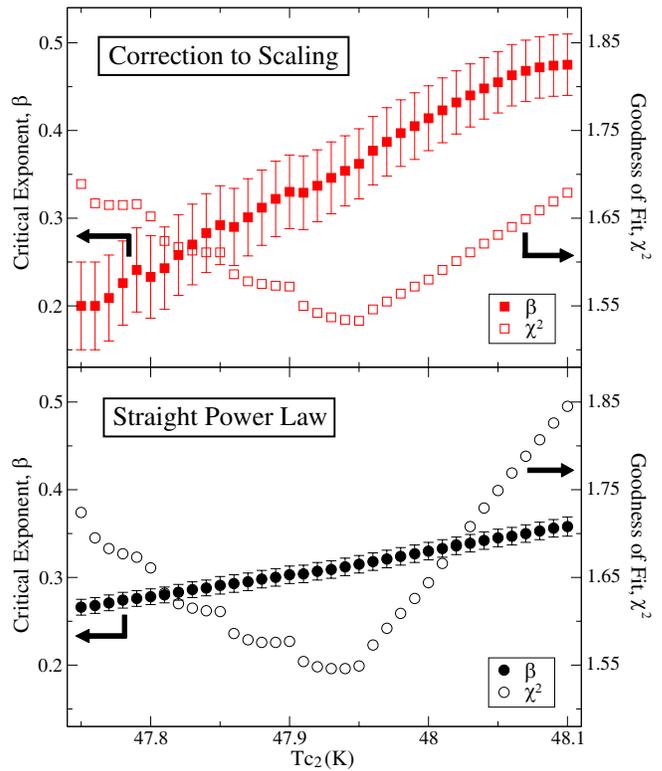}
\caption{(Color online) The dependence of the critical exponent $\beta$ and goodness of fit parameter $\chi^{2}$ on the transition temperature T$_{C2}$.  The bottom panel contains fit parameters obtained using a straight power law fit function (as described by Eq. 1), while the top panel shows parameters obtained from fits to a power law which includes the first order correction-to-scaling term (as described by Eq. 2).  Both sets of fits were performed using data collected above T$_{i}$ = 43 K. }
\end{figure}

One of the factors that complicates the accurate determination of critical exponents is the uncertainty inherent in simultaneously fitting both the critical exponent $\beta$ and the transition temperature T$_{C2}$.  As shown in Fig. 4, these two parameters are intimately related, and the optimal value of $\beta$ is strongly dependent on the appropriate choice of T$_{C2}$.  This situation can be even further complicated by the presence of critical fluctuations, which are typically found in the vicinity of a continuous phase transition and can obscure the precise value of the transition temperature.  Fortunately, as in the case of CuGeO$_{3}$\cite{Lumsden_pure}, the critical fluctuations around T$_{C2}$ in TiOBr are quite difficult to observe.  This is illustrated by the data provided in Fig. 3, which clearly shows a sharp drop in scattering intensity at T$_{C2}$ with very little broadening or rounding near the transition.  This makes TiOBr an excellent candidate for studies of critical behavior, and should allow for a particularly accurate determination of $\beta$ and T$_{C2}$.

\begin{figure}
\includegraphics{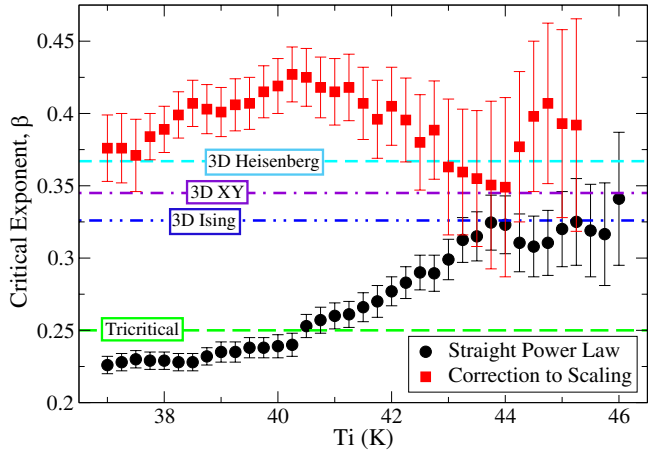}
\caption{(Color online) The value of the critical exponent $\beta$ as a function of T$_{i}$, the start of the fitting range.  The black circles represent values obtained from fits to a straight power law (Eq. 1), while the red squares represent values obtained from fits to a power law which includes the first order correction-to-scaling term (Eq. 2).  The predicted values of $\beta$ for the tricritical and conventional 3D universality classes have been provided as a guide to the eye. }
\end{figure}

The importance of the correction-to-scaling term is illustrated by the data provided in Fig. 5, which shows the effect of fitting range on the value of the critical exponents extracted from Eqs. 1 and 2.  Here the fitting range is described in terms of the initial temperature, T$_{i}$, which represents the start of the fitting range and the lower temperature bound for the data points included in the fit.  Since the upper bound for each fitting range was fixed at a final temperature of T$_{f}$ = 51 K, it follows that T$_{i}$ can be taken as both a measure of the size of a given fitting range and an indication of how far a particular data set extends below the transition temperature T$_{C2}$.  Thus higher values of T$_{i}$ correspond to smaller fitting ranges, and describe fits to data which falls within (or at least close to) the asymptotic critical regime.
  
The exponents determined from fits to the straight power law of Eq. 1 are shown by the black circles in Fig. 5.  Note that these exponents follow a clear trend, gradually increasing as T$_{i}$ rises and the fitting range is reduced.  Closer to the transition temperature (i.e. above T$_{i}$ $\sim$ 43 K) the values of $\beta$ appear to level off, approaching an asymptotic limit of $\beta_{asy}$ = 0.32 $\pm$ 0.03.  This behavior suggests that the asymptotic critical regime in TiOBr is quite narrow, extending less than 5 K (or $\sim$ 9\%) below T$_{C2}$.  Within the critical region, where the critical behavior of the system can accurately be described by a standard power law, Eq. 1 provides consistent and reliable $\beta$ values.  Over larger fitting ranges, for data sets which extend beyond the critical region, deviations from simple power law behavior become increasingly significant and fits to Eq. 1 yield progressively worse values of $\beta$.  Hence $\beta_{asy}$ represents the most accurate value of $\beta$ that can be obtained from Eq. 1, taken in the appropriate asymptotic limit close to T$_{C2}$.

The critical exponents obtained from fits to the modified power law of Eq. 2 are illustrated by the red squares in Fig. 5.  In contrast to the results of the straight power law fits, the values of $\beta$ obtained from fits to Eq. 2 follow a relatively constant trend, with slow variation around $\beta_{avg}$ = 0.39 $\pm$ 0.05.  Thus we find that the introduction of the correction-to-scaling term compensates for deviations from standard power law behavior beyond the asymptotic critical regime, dramatically reducing the effect of fitting range on the resulting critical exponents.  Ideally, one might expect that sufficiently close to T$_{C2}$ the values of $\beta$ from the two fit functions should converge.  In practice, the quality and reliability of fits performed when T$_{i}$ is within $\sim$ 2 K of the phase transition are somewhat limited by the finite size of the data sets, which grow progressively smaller as T$_{i}$ increases.  However, bearing this restriction in mind, it is very encouraging to note that the values obtained for $\beta_{asy}$ from Eq. 1 and $\beta_{avg}$ from Eq. 2 are fully consistent within the bounds of experimental uncertainty.

From these results, we can conclude that the critical behavior of TiOBr can be described by one of the conventional 3D universality classes - 3D Ising ($\beta$ = 0.326), 3D XY ($\beta$ = 0.345) or 3D Heisenberg ($\beta$ = 0.367).  Both $\beta_{asy}$, the result obtained from straight power law fits performed in the appropriate limit, and $\beta_{avg}$, the result obtained from modified power law fits which included a correction-to-scaling term, are consistent with these values.  Due to the relatively large size of our error bars, it is difficult to narrow down the list of potential universality classes to suggest a specific symmetry for the order parameter.  One might only note that the best fits to the data are provided by the modified power law of Eq. 2, which leads to a critical exponent $\beta_{avg}$ indicative of a 3D universality class with continuous symmetry, either XY or Heisenberg.  This would be consistent with previous findings for conventional spin-Peierls systems such as CuGeO$_{3}$\cite{Lumsden_pure} and MEM-(TCNQ)$_{2}$\cite{Lumsden_MEM}.

It should be noted that the success of our analysis of the critical behavior in TiOBr provides extremely compelling evidence that the phase transition at T$_{C2}$ is continuous, rather than first order, in nature.  The fact that the integrated intensity of the superlattice peaks can be fit so well using a continuous model strongly contradicts the recent claims of Schonleber et al\cite{Schonleber_crackpot}, while lending support to earlier findings based on magnetic susceptibility\cite{Seidel, Ruckamp}, NMR\cite{Imai}, specific heat\cite{Hemberger, Ruckamp}, thermal expansion\cite{Ruckamp}, and x-ray diffraction measurements\cite{Shaz, Smaalen, Abel, Sasaki_condmat}.  There can be no confusing the results of the present study with the previously reported critical exponents of $\beta$ = 0.10 to 0.26\cite{Sasaki_JPSJ}, 0.15\cite{Abel}, and 0.25\cite{Smaalen} for TiOBr and TiOCl.  However, our examination of the relationship between the fitting range employed and the resulting value of $\beta$ provides a simple and straightforward explanation for how these anomalously low critical exponents may have come about - they are the result of fits performed to measurements collected outside the asymptotic critical regime.

\begin{figure}
\includegraphics{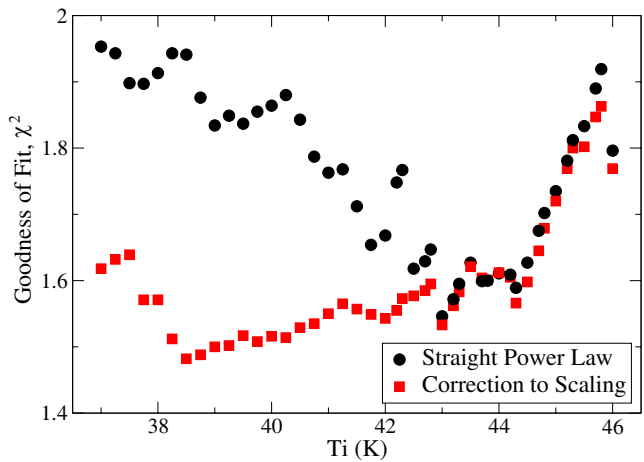}
\caption{(Color online) The goodness of fit parameter $\chi^{2}$ as a function of T$_{i}$, the start of the fitting range.  The black circles correspond to fits performed to a straight power law fit function (Eq. 1), while the red squares correspond to fits to a power law which includes the first order correction-to-scaling term (Eq. 2). }
\end{figure}

The goodness of fit parameter, $\chi^{2}$, is plotted as a function of fitting range in Fig. 6.  Similar to Fig. 5, this plot serves to emphasize the importance of considering correction-to-scaling terms when dealing with data sets which extend beyond the asymptotic critical regime.  The most striking difference between the values of $\chi^{2}$ obtained from Eqs. 1 and 2 emerges over large fitting ranges (i.e. for low values of T$_{i}$).  For T$_{i}$ less than 43 K, as the fitting range grows larger one can observe a gradual monotonic increase in the $\chi^{2}$ values from the straight power law fits.  Again, this reflects the fact that outside the asymptotic critical regime a simple power law form is no longer sufficient to describe the behavior of the order parameter.  In contrast, the $\chi^{2}$ values from fits to the modified power law which incorporates the first order correction-to-scaling term remain small, and relatively constant, down to the lowest values of T$_{i}$.  For small fitting ranges (i.e. for T$_{i}$ greater than 44 K), the quality of fit decreases markedly for both fit functions.  This is simply a result of the reduced number of data points available for fits at higher T$_{i}$, which ultimately becomes the limiting factor for the accuracy of the fits in this regime.  More generally, it is important to note that for any value of T$_{i}$ the goodness of fit is always better when the data is fit to Eq. 2 rather than Eq. 1.  In other words, the correction-to-scaling fits consistently offer a better description of the data than a straight power law, regardless of the fitting range chosen. 

\subsection{Implications for TiOCl}

\begin{figure}
\includegraphics{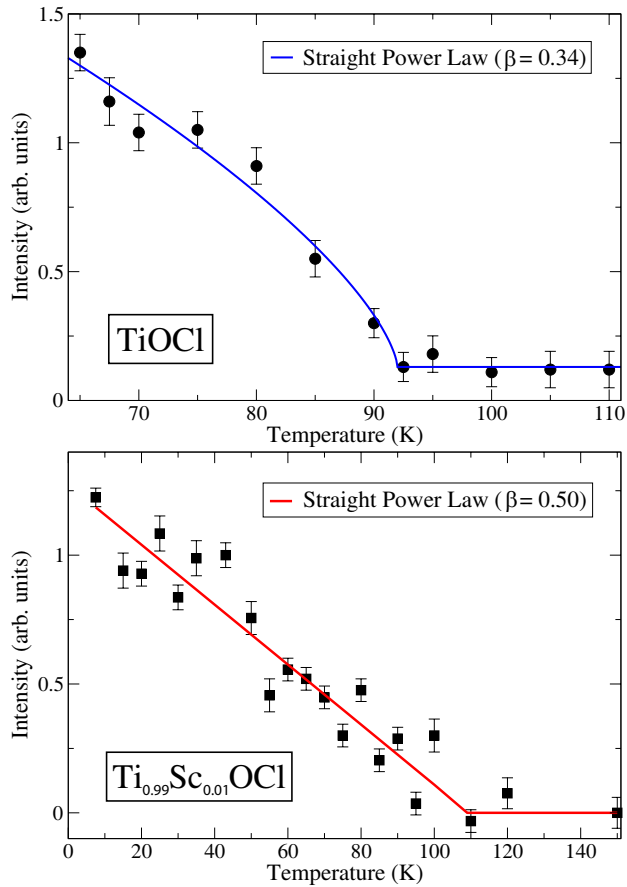}
\caption{(Color online) The temperature dependence of the integrated scattering intensity at the (2 $\pm$ $\delta_{H}$, 1.5, 1) superlattice peak position in TiOCl (top) and Ti$_{0.99}$Sc$_{0.01}$OCl (bottom).  These intensities have been obtained by integrating over H-scans of the form shown in the colour contour maps of figure 1.  The TiOCl data set can be well described by a straight power law fit which is consistent with conventional 3D universality ($\beta$ = 0.34, shown in blue), while the Ti$_{0.99}$Sc$_{0.01}$OCl data set is much better fit by a power law function consistent with mean-field-like behavior ($\beta$ = 0.50, shown in red).
%The fit functions shown in the upper panel correspond to a straight power law with $\beta_{asy}$ = 0.32 (red) and a power law including the first correction-to-scaling term with $\beta_{avg}$ = 0.39 (blue).  The fit function in the bottom panel corresponds to a straight power law with $\beta$ = 0.50 (red).%
}
\end{figure}

Having completed our analysis of the critical properties of TiOBr, it is natural that we next turn our attention to what our results can suggest about the critical behavior of TiOCl.  While we lack sufficient data to perform the analysis necessary to precisely determine the value of $\beta$ for TiOCl, we can clearly demonstrate that the critical exponents determined in the previous section provide a very convincing description of recently reported measurements on TiOCl\cite{Clancy_pure}. 

As illustrated by the colour contour maps provided in Fig. 1, the low temperature phase diagram of TiOCl is considerably more complicated than that of TiOBr.  In particular, TiOCl is distinguished by the presence of commensurate fluctuations which persist from T$_{C1}$ up to T*, throughout both the incommensurate spin-Peierls and pseudogap phases.  However, since the order parameter associated with the transition at T$_{C2}$ is strictly related to the intensity of the incommensurate satellite peaks, rather than the sum of all dimerization scattering, it is only the incommensurate scattering at (H $\pm$ $\delta_{H}$, K+1/2, L) which must be considered when studying the critical behavior of TiOCl.
  
The upper panel of Fig. 7 shows the temperature dependence of the integrated scattering intensity of the (2 $\pm$ $\delta_{H}$, 1.5, 1) incommensurate satellite peaks in TiOCl.  These intensities were obtained by integrating over small {\bf Q}-ranges around the incommensurate ordering wave vectors in H-scans similar to those used to construct the colour contour maps of Fig. 1.  The data were fit to a standard power law, as described by Eq. 1, although due to the low density of points near T$_{C2}$ the value of the transition temperature was fixed at 92 K.  This fixed value of T$_{C2}$ is consistent with previously reported susceptibility\cite{Seidel, Ruckamp}, heat capacity\cite{Hemberger, Ruckamp}, NMR\cite{Imai}, and x-ray diffraction measurements\cite{Krimmel, Abel, Clancy_pure} for TiOCl.  The resulting line of best fit (indicated by the solid blue line in the upper panel of Fig. 7) corresponds to a critical exponent of $\beta$ = 0.34 $\pm$ 0.04.  It should be noted that this value of $\beta$ is consistent with the theoretical predictions of the 3D Ising, 3D XY and 3D Heisenberg models, indicating that the transition in TiOCl can be well described by any of the conventional 3D universality classes.  Furthermore, there is excellent agreement between the value of $\beta$ = 0.34 $\pm$ 0.04 for TiOCl and the value of $\beta_{asy}$ = 0.32 $\pm$ 0.03 obtained earlier from power law fits to the TiOBr data.  In fact, this relatively simple estimate for $\beta$ also agrees quite well with the value of $\beta_{avg}$ = 0.39 $\pm$ 0.05 obtained for TiOBr using the modified power law of Eq. 2, in spite of the fact that the large fitting range used for the TiOCl data set almost certainly extends well beyond the asymptotical critical regime.

Unfortunately, due to the small size and low point density of the TiOCl data set, it is difficult to extract any meaningful parameters from fits performed using the modified power law described by Eq. 2.  On a more qualitative level however, it is worth noting that the function in Eq. 2 can fit the TiOCl data very well using a fixed transition temperature of T$_{C2}$ = 92 K and the critical exponent of $\beta_{avg}$ = 0.39 determined for TiOBr.  By eye, such a fit is almost indistinguishable from the solid blue curve which represents the best fit line for Eq. 1, hence it has been omitted from the upper panel of Fig. 7.
  
A very interesting comparison can also be made using previous x-ray data collected on Ti$_{0.99}$Sc$_{0.01}$OCl, a compound in which Sc-doping has been shown to suppress commensurate spin-Peierls order to below at least T $\sim$ 7 K\cite{Clancy_doped}.  The temperature dependence of the integrated intensity of the incommensurate scattering at (2 $\pm$ $\delta_{H}$, 1.5, 1) in Ti$_{0.99}$Sc$_{0.01}$OCl is provided in the bottom panel of Fig. 7.  If we attempt to fit this data using the simple power law fit function described by Eq. 1 we find that the critical behavior of Sc-doped TiOCl is clearly different from that of either pure TiOCl or TiOBr.  The best power law fit to the Ti$_{0.99}$Sc$_{0.01}$OCl data (illustrated by the solid red line in the bottom panel of Fig. 7) corresponds to a critical exponent of $\beta$ = 0.50 $\pm$ 0.07 and a transition temperature of T$_{C2}$ = 109 $\pm$ 6 K.  This value of $\beta$ is consistent with the exponent predicted by mean field theory ($\beta$ = 0.5), and is incompatible with the theoretical predictions for any of the conventional 3D universality classes.  In addition, the value of $\beta$ for Ti$_{0.99}$Sc$_{0.01}$OCl is dramatically different from the critical exponents that we determined for TiOCl and TiOBr.  This difference in critical behavior is evident even from the raw data provided in Fig. 7.  The intensity of the incommensurate scattering in Ti$_{0.99}$Sc$_{0.01}$OCl appears to grow linearly with decreasing temperature, while the scattering in TiOCl clearly follows a sublinear power law trend.

It is also instructive to compare these results to previous studies on the critical properties of CuGeO$_3$\cite{Lumsden_pure, Lumsden_doped}.  It has been well established that in both pure CuGeO$_3$ and CuGeO$_3$ doped with ions of similar ionic radius the critical behavior of the spin-Peierls transition can be described by one of the conventional 3D universality classes with continuous symmetry (such as 3D XY or 3D Heisenberg).  However, when CuGeO$_3$ is doped with ions of larger ionic radius, by substituting Cd$^{2+}$ for Cu$^{2+}$ or Ga$^{4+}$ for Ge$^{4+}$ for example, the observed universality class is found to change towards mean-field-like behavior.  This effect appears to have little or no concentration dependence, and is believed to be related to the presence of local strain fields induced by the larger dopant ions.  The significance of local strain effects on the behavior of the order parameter is reasonably intuitive given both the magneto-elastic nature of the spin-Peierls transition and the underlying importance of lattice properties in the spin-Peierls materials.  Hence, if the dopant ion is smaller or of comparable size to the ion it is replacing (i.e. if $\Delta_{ion}$ is less than roughly +3\% of the ionic radius) the universality class of the transition remains unchanged.  Alternatively, for large, positive changes in ionic radius (i.e. $\Delta_{ion}$ is on the order of +9\% or more) there appears to be a distinct change in critical properties.  The difference in ionic radius between Sc$^{3+}$ and Ti$^{3+}$ corresponds to $\Delta_{ion}$ = +11\%.  Thus, the change in critical behavior we observe between TiOCl and Ti$_{0.99}$Sc$_{0.01}$OCl appears fully consistent with lessons learned from studies of doped CuGeO$_{3}$\cite{Lumsden_doped}.  Although further study is still required to provide a more detailed understanding of this phenomenon, on the basis of these results it is certainly tempting to conclude that doping-induced changes in criticality are a fundamental property associated with the spin-Peierls transition.

A second significant feature which arises from the comparison of the TiOCl and Ti$_{0.99}$Sc$_{0.01}$OCl data sets is the change in transition temperature, T$_{C2}$, which is induced by Sc-doping.  As shown by Fig. 7, the value of T$_{C2}$ is considerably higher in Ti$_{0.99}$Sc$_{0.01}$OCl than in TiOCl, with the fitted value of T$_{C2}$ = 109 $\pm$ 6 K for the doped sample lying well above the accepted value of 92 $\pm$ 1 K for the pure sample.  This result is quite surprising, particularly when one considers that doping has been universally found to suppress the spin-Peierls transition temperature in CuGeO$_{3}$.  One possible explanation for this effect is that since no commensurate fluctuations are observed below T* $\sim$ 135 K in Ti$_{0.99}$Sc$_{0.01}$OCl, the T $\sim$ 109 K temperature scale actually corresponds to the appropriately lowered equivalent of T* from TiOCl.  This argument is supported by the fact that the incommensurate scattering in Ti$_{0.99}$Sc$_{0.01}$OCl is very broad, implying that the length scale for correlations in the doped compound is even shorter than that of the commensurate fluctuations in pure TiOCl.  In addition, the $\sim$ 20 \% drop between the 135 K and 109 K temperature scales is quite consistent with the 15 to 45 \% suppression of T$_{SP}$ per 1 \% concentration of dopants which has been reported in the literature for doped CuGeO$_{3}$\cite{Hase, Oseroff, Schoeffel, Manabe, Martin, Grenier, Anderson, Mgdoped}.

\subsection{Details of the Incommensurate Structure of TiOBr}

As discussed earlier, there are several respects in which the low temperature scattering of TiOBr differs from that of its isostructural counterpart TiOCl.  The chief feature which distinguishes these two compounds from each other is the presence of commensurate fluctuations, which are clearly evident in TiOCl\cite{Clancy_pure} but appear to be absent in TiOBr.  In TiOCl, these commensurate fluctuations can be observed throughout both the incommensurate spin-Peierls phase (from T$_{C1}$ to T$_{C2}$) and the pseudogap phase (from T$_{C2}$ to T*), as shown in the right panel of Fig. 1.  The commensurate scattering is not resolution-limited above T$_{C1}$, indicating that it does not correspond to static long-range order.  However, the width of the scattering implies the presence of relatively long correlation lengths, on the order of $\sim$ 100 {\AA} or more.  Interestingly, between T$_{C1}$ and T$_{C2}$ the commensurate fluctuations appear to both coexist with and compete against the incommensurate scattering at (H $\pm$ $\delta_{H}$, K + 1/2, L).  The commensurate fluctuations also appear to be shifted in {\bf Q}-space with respect to both the accompanying incommensurate scattering and the long-range ordered commensurate state below T$_{C1}$, with observed peak positions displaced along both the H and K directions.

In TiOBr, we find no indication of commensurate fluctuations in either the incommensurate spin-Peierls state (between T$_{C1}$ and T$_{C2}$) or the pseudogap state (between T$_{C2}$ and T*).  This was the case at each of the 7 different commensurate ordering wave-vectors investigated in this study.  One possible explanation for the absence of commensurate fluctuations in TiOBr may arise from the subtle differences between the structures of TiOCl and TiOBr.  While the distance between neighboring dimerized chains of Ti atoms (i.e. the interchain spacing, related to the \textit{a} lattice parameter) is almost identical in both two compounds, the distance between neighbouring Ti atoms on the same dimerized chain (i.e. the intrachain spacing, given by the \textit{b} lattice parameter) is approximately 4 \% larger in TiOBr.  As a result of this change in lattice parameters, TiOBr appears to be less one-dimensional than its isostructural counterpart.  The difference is clearly evident from the magnetic susceptibility measurements reported for the two compounds\cite{Seidel, Sasaki_JPSJ, Kato}, both in terms of the magnitude of the nearest-neighbor exchange coupling (which is estimated to be $\sim$ 45 \% smaller in TiOBr) and the observed deviations from the idealized Bonner-Fisher curve (which suggest significantly stronger interchain couplings in TiOBr).  Doping studies on TiOCl\cite{Clancy_doped} indicate that the relative strengths of the intrachain and interchain interactions are intimately connected with the stability of the incommensurate structure.  Thus, if the interchain coupling is stronger with respect to the intrachain coupling in TiOBr than it is in TiOCl then it could easily lead to suppression of commensurate fluctuations (as in Ti$_{0.99}$Sc$_{0.01}$OCl), stabilization of the incommensurate structure, and a loss of phase coexistence between T$_{C1}$ and T$_{C2}$.

It should be emphasized that the lack of observable commensurate fluctuations in the present study does not necessarily imply that such scattering is completely absent in TiOBr.  In fact, the difficulty in detecting such a weak, broad feature - particularly one which may be slightly shifted in {\bf Q}-space - is clearly underscored by the number of x-ray diffraction studies which previously failed to observe such commensurate fluctuations in TiOCl.  It should also be noted that the (2, 1.5, 1) wave-vector, where the commensurate scattering was found to be strongest in TiOCl, was experimentally inaccessible in this study.  Thus, there remains a possibility that stronger commensurate fluctuations could still be found in TiOBr at other, as of yet unstudied, positions in reciprocal space.  Further work is needed to provide a more thorough understanding of this intriguing difference between the TiOX spin-Peierls systems.

\begin{figure}
\includegraphics{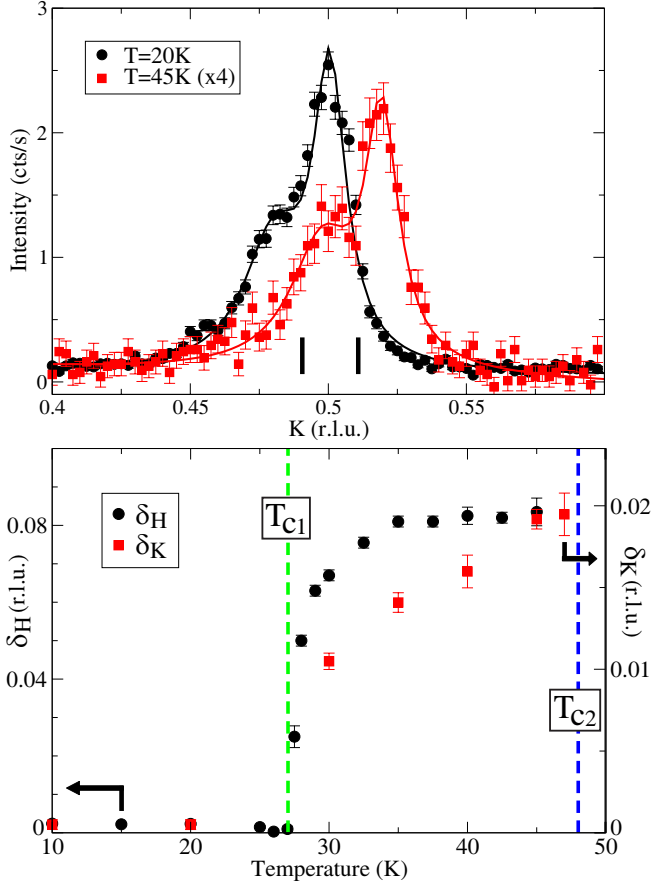}
\caption{(Color online) Representative K-scans performed through the (-$\delta_{H}$, 0.5, 6) superlattice peak in TiOBr (top).  The black circles correspond to data collected within the commensurate spin-Peierls state (T = 20 K, $\delta_{H}$ = 0), while the red squares correspond to data collected within the incommensurate spin-Peierls state (T = 45 K, $\delta_{H}$ $\sim$ 0.08).  For illustrative purposes the intensity of the T = 45 K data set has been scaled by a factor of four.  The solid lines represent fits to a simple two-Gaussian function.  Note that the peaks in the incommensurate phase are shifted in {\bf Q}-space relative to those of the commensurate phase.  The temperature dependence of the shift, $\delta_{K}$, is shown in the bottom panel, along with the H-component of the incommensurate modulation wave vector, $\delta_{H}$. }
\end{figure}

Our measurements in the incommensurate spin-Peierls phase also reveal an interesting property of the incommensurate modulation in TiOBr.  In addition to the primary component of the incommensurate modulation wave-vector which runs along the H direction (perpendicular to the direction of the dimerized chains), a smaller secondary component of the modulation has been reported along the K direction (parallel to the dimerized chains) in both TiOCl and TiOBr.  The magnitude of the K-component of the incommensurate modulation, $\delta_{K}$, increases monotonically between T$_{C1}$ and T$_{C2}$, reaching a maximum value of $\sim$ 0.015 r.l.u. in both compounds\cite{Sasaki_condmat, Abel}.  Even though the experimental resolution of our present study is too broad to fully resolve the distinct incommensurate satellite peaks which should arise from the K-component of the incommensurate modulation, we would anticipate that the effects of this secondary incommensuration should be manifested as a significant broadening of the superlattice peaks along the K-direction.  However, as the representative K-scans provided in the upper panel of Fig. 8 clearly demonstrate, there is no observable change in the width of the (-$\delta_{H}$, 0.5, 6) superlattice peak between the commensurate and incommensurate spin-Peierls phases.  What these K-scans do reveal is an obvious shift in the position of the incommensurate satellite peaks above T$_{C1}$ relative to the commensurate superlattice peaks below T$_{C1}$.  The temperature dependence of this shift in {\bf Q}-space, which we will call $\delta_{K}$, is illustrated in the bottom panel of Fig. 8.

Note that superlattice reflections are observed at the ($\pm$ $\delta_{H}$, 0.5 + $\delta_{K}$, 6) positions, but not at ($\pm$ $\delta_{H}$, 0.5 - $\delta_{K}$, 6).  In this respect, our K-shift clearly differs from the previously reported K-incommensuration, which when coupled with the incommensuration along H should produce a set of 4 incommensurate peaks at ($\pm$ $\delta_{H}$, 0.5 $\pm$ $\delta_{K}$, 6).  Furthermore, there are indications that this K-shift exhibits some form of systematic {\bf Q}-dependence.  At the (0, 0.5, 6), (0, -0.5, 6), and (1, 0.5, 6) positions in reciprocal space we find the incommensurate peaks shift to higher {\bf Q} positions (i.e. to K + $\delta_K$), while for (0, 0.5, 5) and (1, 0.5, 5) we find the peaks shift to lower Q (i.e. to K - $\delta_K$).  At the (0, 1.5, 6) wave-vector no discernable K-shift could be observed.  Based on alignment scans performed through several nearby structural Bragg peaks we can conclude that this shift in {\bf Q}-space is not simply the result of a change in the lattice constants of the material.  Interestingly, the shift in {\bf Q}-space appears to have a different temperature dependence than the H-component of the incommensurate modulation wave-vector, as shown in the bottom panel of Fig. 8.  While $\delta_{H}$ and $\delta_{K}$ both increase steadily with temperature above T$_{C1}$, $\delta_{H}$ approaches saturation by $\sim$ 35 K, while $\delta_{K}$ continues to grow until the system reaches T$_{C2}$.  If the observed K-shift was simply equivalent to the K-component of the incommensurate modulation wave-vector it seems surprising that it would not follow the same temperature evolution as $\delta{H}$. 

This effect has also been observed in recent x-ray measurements performed by Sasaki et al.\cite{Sasaki_condmat}, which show that the incommensurate scattering at ($\pm$ $\delta_{H}$, 2.5, 0) is shifted in the K-direction (towards lower {\bf Q}) with respect to the commensurate scattering at (0, 2.5, 0).  Furthermore, the magnitude and temperature dependence of this shift in {\bf Q}-space are almost identical to that reported here.  Curiously, in spite of the absence of additional satellite peaks at ($\pm$ $\delta_{H}$, 2.5 + $\delta_{K}$, 0), Sasaki et al. treat this behavior as equivalent to the K-incommensuration reported by van Smaalen et al.\cite{Smaalen}.  This assumption is not completely unjustified, particularly since recent x-ray diffraction measurements on TiOCl\cite{Abel} have shown that in some cases there can be dramatic differences between the observed scattering intensities of neighbouring pairs of incommensurate satellite peaks.  These intensity variations have been attributed to changes in structure factor resulting from phase shifts between the modulations of neighboring chains of Ti atoms in the unit cell.  In TiOCl these intensity differences have been found to be up to one to two orders of magnitude in size\cite{Abel}.  If the scattering at the ($\pm$ $\delta_{H}$, 0.5 - $\delta_{K}$, 6) position in TiOBr is suppressed by a similar factor then it could be too weak to be detectable with our current experimental configuration.  However, while it is possible that weak superlattice scattering could be obscured by experimental background in the present study, it is less likely that this same problem would also have affected the results of Sasaki et al. due to the much higher signal-to-noise ratio available at synchrotron x-ray sources.  

An alternative possibility, which naturally arises from the parallels between TiOCl and TiOBr, is that the K-shift between the commensurate and incommensurate scattering in TiOBr is an analog of the {\bf Q}-shift between the commensurate fluctuations and long-range ordered commensurate/incommensurate scattering in TiOCl.  The magnitude of the {\bf Q}-shift reported for the commensurate fluctuations in TiOCl\cite{Clancy_pure} is very similar to the size of the K-shift we observe in TiOBr, and it is clear that both of these features are intimately connected with the first-order phase transition at T$_{C1}$.  Thus, the shift in {\bf Q}-space between the commensurate and incommensurate scattering in TiOBr could represent a distinctly different effect from the K-component of the incommensurate modulation.  In this case, the K-shift could reflect changes in the balance of interchain and intrachain interactions at T$_{C1}$.  Hence, below and above T$_{C1}$ the ratio of relevant interaction strengths favors commensurate and incommensurate modulations of the dimerized structure with two slightly different wave-vectors.  It is difficult to comment on the specific implications of these {\bf Q}-shifts without further, and more thorough, study of additional superlattice peak positions.  However, this result may provide an indication that the structure of the incommensurate spin-Peierls state of TiOBr may be more complex than previously assumed.

\section{Conclusions}

In conclusion, we have performed the first detailed study of the critical properties of TiOBr.  While the spin-Peierls behavior of this system appears to be unconventional in a number of respects, the critical properties of the system can be very well described by conventional 3D universality classes.  This finding is consistent with previous results obtained from studies performed on other spin-Peierls systems such as MEM-(TCNQ)$_{2}$\cite{Lumsden_MEM} and CuGeO$_{3}$\cite{Lumsden_pure}.  By analyzing our data using a simple power law fit function we determined that the critical exponent associated with the transition at T$_{C2}$ approaches a value of $\beta_{asy}$ = 0.32 $\pm$ 0.03 in the appropriate asymptotic critical regime.  The strong dependence of $\beta$ on the size of the fitting range suggests that the asymptotic critical region in TiOBr is quite narrow, as in CuGeO$_{3}$\cite{Lumsden_pure}.  This provides an explanation for the apparently contradictory critical exponents that have previously been reported for this compound.  The narrowness of the critical regime can be accounted for by employing a modified power law fit function which includes the first order correction-to-scaling confluent singularity term.  Using this approach we obtain a value of $\beta_{avg}$ = 0.39 $\pm$ 0.05, a result which is consistent with both the 3D XY and 3D Heisenberg universality classes.

Based on a simple, primarily qualitative comparison, we argue that TiOCl, the isostructural cousin of TiOBr, can also be well described by one of the conventional 3D universality classes.  More detailed studies would still be required in order to offer a more quantitative description of the critical behavior of TiOCl, or to further narrow down the list of potential universality classes for the transition.  However, this result appears to add growing weight to the notion that the spin-Peierls transition is by nature a conventional three-dimensional phase transition, whether it is realized in organic or inorganic systems or even the ``unconventional" titanium oxyhalides.  Applying the same approach to measurements performed on Ti$_{0.99}$Sc$_{0.01}$OCl we find evidence that Sc-doping induces a fundamental change in the critical properties of the transition at T$_{C2}$.  We suggest that this effect is likely caused by the presence of local lattice strains which arise due to the significantly larger ionic radius of the Sc$^{3+}$ dopant ions.

In addition, we describe several interesting features of the incommensurate structure of TiOBr between T$_{C1}$ and T$_{C2}$.  Our measurements reveal no evidence for commensurate fluctuations of the type observed in the incommensurate spin-Peierls and pseudogap phases of TiOCl.  This result may reflect the weaker one-dimensional properties of TiOBr, and in particular the substantially smaller size of the intrachain interaction.  We also observe that the commensurate and incommensurate structures are shifted in {\bf Q}-space with respect to one another, with a displacement along the K-direction that steadily grows between T$_{C1}$ and T$_{C2}$.  We hope that these measurements will help to guide and inform future theoretical and experimental studies of these novel spin-Peierls systems.

\begin{acknowledgments}

The authors would like to acknowledge helpful discussions with T. Imai, J.P. Castellan, J. Ruff and A. Aczel.  This work was supported by NSERC of Canada and NSC of Taiwan.

\end{acknowledgments}

%
% now the references. delete or change fake bibitem. delete next three
%   lines and directly read in your .bbl file if you use bibtex.
%\begin{thebibliography}{}
%\end{thebibliography}
% figures follow here
%
%
% Here is an example of the general form of a figure:
% Fill in the caption in the braces of the \caption{} command. Put the label
% that you will use with ref{} command in the braces of the \label{} command.
%
%
% Uncomment the following to put the figures into the latex file.
%
%\begin{figure}
%\centering
%\includegraphics[width=8.5cm]{fig4_nolines_tilt.ps}
%\caption{Enter caption here.}
%\label{fig4}
%\end{figure}
%
%\begin{figure}
%\centering
%\includegraphics[angle=0,origin=c,width=8cm]{all4_2.ps}
%\caption{Caption}
%\label{fig2}
%\end{figure}
%
%\begin{figure}
%\centering
%\includegraphics[angle=0,origin=c,width=8cm]{th2thputtogether2.ps}
%\caption{Caption}
%\label{fig3}
%\end{figure}
%
%\begin{figure}
%\centering
%\includegraphics[angle=0,origin=c,width=8cm]{dspace.ps}
%\caption{Caption}
%\label{fig4}
%\end{figure}
%

\end{document}